\documentstyle[epsfig,preprint,aps]{revtex}           
\begin{document}       
\tighten

\def\bea{\begin{eqnarray}}          
\def\eea{\end{eqnarray}}          
\def\beas{\begin{eqnarray*}}          
\def\eeas{\end{eqnarray*}}          
\def\nn{\nonumber}          
\def\ni{\noindent}          
\def\G{\Gamma} 
\def\D{\Delta} 
\def\P{\Pi}        
\def\d{\delta}          
\def\l{\lambda}          
\def\g{\gamma}          
\def\m{\mu}          
\def\n{\nu}
\def\k{\kappa}          
\def\s{\sigma}          
\def\tt{\theta}          
\def\b{\beta}          
\def\a{\alpha}          
\def\f{\phi}          
\def\fh{\hat{\phi}}          
\def\y{\psi}          
\def\z{\zeta}          
\def\p{\pi}          
\def\e{\epsilon}           
\def\ve{\varepsilon}
\def\cd{{\cal D}}          
\def\cl{{\cal L}}          
\def\cv{{\cal V}}          
\def\cz{{\cal Z}}          
\def\pl{\partial}          
\def\ov{\over}          
\def\~{\tilde}          
\def\rar{\rightarrow}          
\def\lar{\leftarrow}          
\def\lrar{\leftrightarrow}          
\def\rra{\longrightarrow}          
\def\lla{\longleftarrow}          
\def\8{\infty} 
\def\pls{\partial\!\!\!/}
\def\bs{b\!\!\!/}
\def\ps{p\!\!\!/}
\def\qs{q\!\!\!/}
\def\ls{l\!\!/}
\def\rs{r\!\!\!/}
\def\ks{k\!\!\!/}
\def\As{A\!\!\!/}
\def\yb{\bar{\y}}
\def\As{A\!\!\!/} 
\def\Cs{C\!\!\!\!/} 
\def\Ds{D\!\!\!\!/} 
\def\Es{E\!\!\!\!/} 
\def\aas{a_1\!\!\!\!/}
\def\bbs{b_{11}\!\!\!\!\!\!/}
\def\bbbs{b_{12}\!\!\!\!\!\!/}
\def\bbbbs{b_{2}\!\!\!\!/}

\title{Radiatively-Induced 
Lorentz and CPT Violating Chern-Simons Term in QED}
          
\author{J.-M. Chung\footnote{Electronic address: chung@ctpa03.mit.edu}}
\address{Center for Theoretical Physics,
Massachusetts Institute of Technology,\\
Cambridge, Massachusetts 02139}
       
\date{MIT-CTP-2864,~~~~ May 1999}                  
\maketitle              
\draft              
\begin{abstract}           
\indent           
We calculate the induced Lorentz- and CPT-violating Chern-Simons term 
arising from the Lorentz- and CPT-violating sector of quantum electrodynamics
with a $b_\m\bar{\y}\g^\m\g_5\y$ term. The result to all orders in $b$ 
coincides with the previous linear-in-$b$ calulation by Chung and Oh 
[hep-th/9812132] as well as Jackiw and 
Kosteleck\'{y} [Phys. Rev. Lett. {\bf 82}, 3572 (1999)], since all higher
order terms vanish. 
\end{abstract}              
                   
Recently, Colladay and Kosteleck\'{y} \cite{ck} posed the question whether
a  Lorentz- and CPT-violating
Chern-Simons term \cite{djt} is induced  when the Lorentz- and CPT-violating term
$\bar{\y}\bs\g_5\y$ ($b_\m$ a constant 4-vector) is added to the 
conventional Lagrangian of QED \cite{cfj}. This has been 
discussed by a number of authors \cite{co,jk,ch,jm,fg,pv}.
In Refs.~\cite{ck,co,jk,ch,jm}, only the term linear in $b_\m$ was kept in
the calculation of the vacuum polarization (which determines the induced 
Chern-Simons term).
A calculation of higher order (in $b$) corrections was attempted 
by Fosco and Le Guillou \cite{fg}. However, their non-trivial result 
is invalid because they use 
an incorrect fermion propagator. After completion of our exact, i.e., 
non-perturbative in $b_\m$, calculation,  
we received a paper by P\'{e}rez-Victoria \cite{pv},    
in which the same calculation was carried out, with the
same result as we report  here: for $b<m$ all higher order terms vanish. 

Consider a Dirac fermion propagating in a Lorentz- 
and CPT-violating manner in the background of a photon field. 
The Lagrangian of this system is 
\bea
\cl=i\bar{\y}\pls\y-m\bar{\y}\y-\bar{\y}\bs\g_5\y-\bar{\y}\As\y\;,
\label{lg}
\eea
where $b_\m$ is a constant 4-vector, and $b_\m\bar{\y}\g^\m\g_5\y$
is the Lorentz violating, CPT-odd term.  
The fermion propagator is 
\bea
G(p)={i\ov \ps-m-\bs\g_5}\;.\label{pg}
\eea
If we observe the following relation
\beas
&&(\ps-m-\bs\g_5)(\ps+m-\bs\g_5)(\ps+m+\bs\g_5)(\ps-m+\bs\g_5)\nn\\
&&~~~~~~~~~~~~~~~~~~~~~~~~~~~~~~~~~~~~~~~~~~~~~
=(p^2-m^2-b^2)^2-4(p\!\cdot\! b)^2+4p^2b^2\;,
\eeas
we can readily rationalize the propagator, Eq.~(\ref{pg}), as follows:
\bea
G={i\ov \ps-m-\bs\g_5}= {i(U+V\g_5)\ov D}\;,
\eea
where
\beas
&&U=(p^2-m^2+b^2)(\ps+m)-2mb^2-2p\!\cdot\! b \bs\;,\nn\\
&&V=(p^2-m^2+b^2)\bs-(2p\!\cdot\! b+2m\bs)(\ps-m)\;,\nn\\
&&D=(p^2-m^2-b^2)^2-4(p\!\cdot\! b)^2+4p^2b^2\;.
\eeas

The vacuum polarization tensor is given by
\bea
\P^{\m\n}(p)=\int {d^4k\ov (2\p)^4}{\rm tr}[\g^\m G(p+k)\g^\n G(k)]\;,
\label{vpt}
\eea
and its antisymmetric part has the following general structure:
\bea
\P^{[\m\n]}&\equiv& {1\ov 2}(\P^{\m\n}-\P^{\n\m})\nn\\
&=&\e^{\m\n\a\b}p_\a b_\b \P(p^2,p\!\cdot\! b, b^2)
+(p^\m b^\n-b^\m p^\n)F(p^2,p\!\cdot\! b, b^2)\;.
\eea
The derivative of $\P^{[\m\n]}$ with respect to the external momentum
$p_\a$ at $p_\a=0$ gives
\bea
{\pl\ov \pl p_\a}\P^{[\m\n]}\bigg|_{p=0}&=&
\e^{\m\n\a\b}b_\b \P(0, 0, b^2)
+(g^{\m\a}b^\n-g^{\n\a}b^\m)F(0, 0, b^2)\;.\label{p0}
\eea
The first term on the right-hand side of the above equation
produces the induced Chern-Simons term in the effective action as follows:
\bea
{1\ov 2}\e^{\m\n\a\b}k_\m A_\n F_{\a\b}\;,~~~{\rm with}~~~
k_\m=-{1\ov 2}b_\m \P(0,0,b^2)\;.\label{cs}
\eea
Since our concern here is to obtain the induced Chern-Simon term,
we do not consider the second term on the right-hand side
of Eq.~(\ref{p0}); we only collect terms propotional to $\e^{\m\n\a\b}$ from
${\pl\ov \pl p_\a}\P^{[\m\n]}|_{p=0}$.
The complicated trace algebra contains  
eight gamma matrices together with $\g_5$. We perform these trace calculations
using the package ``Feynpar.m'' which runs in the Mathematica System. 
(If interested readers request the trace calculation program, 
the author will gladly provide it.) 
After the trace calculation, we are left with the momentum integral: 
\bea
\e^{\m\n\a\b}b_\b \P(0, 0, b^2)=-4i\e^{\m\n\a\b}b_\b
\int{d^4 k\ov (2\p)^4}{R_1+(k\!\cdot\! b)^2 R_2+(k\!\cdot\! b)^4 R_3\ov 
D^3}\;,\label{kint}
\eea
where
\beas
R_1&=&-b^8 - 2 b^6 k^2 + 2 b^2 k^6 + k^8 - 6 b^6 m^2 - 4 b^4 k^2 m^2 +
   2 b^2 k^4 m^2 \nn\\
&-&12 b^4 m^4 + 6 b^2 k^2 m^4 -6 k^4 m^4 -
   10 b^2 m^6 +8 k^2 m^6- 3 m^8 \;,\nn\\
R_2&=&8 b^4 + 4 b^2 k^2 - 8 k^4 - 4 k^6/b^2 + 28 b^2 m^2 - 16 k^2 m^2\nn\\
&+& 12 k^4 m^2/b^2 + 24 m^4 - 12 k^2 m^4/b^2 + 4 m^6/b^2 \;,\nn\\
R_3&=&-16 + 16 k^2/b^2 -16 m^2/b^2 \;.
\eeas
In order to perform this 4-dimensional Minkowski-momentum integral,
we first change the time-component of the $k$ and $b$ vectors, $k^0$ and $b^0$,
into 
\beas
&&k^0\rar k^4=ik^0\;, \nn\\ 
&&b^0\rar b^4=ib^0\;,
\eeas
to get an Eucledian metric. This is allowed as long as $b^2<m^2$. Then we use
4-dimensional spherical polar coordinates.
Furthermore, with replacements $k^2=m^2 x$ and $b^2=m^2 y$,
and an introduction of a parameter $\e$, which is set to zero 
before the radial $k$ (or $x$) integration, 
in the following fashion:
\beas
{1\ov D^3}={1\ov 2}{\pl^2\ov \pl \e^2}{1\ov D+\e}\bigg|_{\e\rar0}\;,
\eeas
the right-hand side of Eq.~(\ref{kint}) becomes
\bea
&&
{\e^{\m\n\a\b}b_\b\ov 4\p^3}{\pl^2\ov\pl\e^2}
\int_0^\8 x dx\int_0^\p \sin^2\tt d\tt\biggl[
{A+B \cos^2\tt+C \cos^4\tt\ov \a\sin^2\tt+\b+\e}\biggr]\bigg|_{\e=0}\nn\\
&=&{\e^{\m\n\a\b}b_\b\ov 2\p^2}\int_0^\8 dx
\biggl\{{Cx\ov \a^3}
+\biggl[{A(\a+4\b)\ov (\a+\b)^2}+{B\ov \a+\b}
+{C(\a^2-4\a\b-8\b^2)\ov \a^3}\biggr]{x\ov 8\b\sqrt{\b(\a+\b)}}\biggr\}\;,
\label{xi}
\eea
where
\beas
A&=&(x-y+1)(x+y-3)(\a+\b)\;,\nn\\
B&=&-4x(x-y+1)(\a+\b+y[x+y-3])\;,\nn\\
C&=&16x^2 y(x-y+1)\;,\nn\\
\a&=&4xy\;,\nn\\
\b&=&(x-y+1)^2\;.
\eeas
The second term in the integrand of Eq.~(\ref{xi}) may be further simplified as
\beas
{1\ov 2}\biggl[1+{x\ov y}-{\a+\b\ov 2y^2}-{(5+y)x\ov \a+\b}
-{(1-y)^2\ov \a+\b}\biggr]{1\ov \sqrt{\a+\b}}\;.
\eeas
The final result of our calculation is
\bea
\P(0,0,b^2)=-{3\ov 8\p^2}\;.\label{fr}
\eea
From Eqs.~(\ref{fr}) and (\ref{cs}), we find that the strength of the induced 
Chern-Simons term is given as
\bea
k_\m={3b_\m\ov 16\p^2}\;.
\eea

This coincides with the 
previous result by Chung and Oh \cite{co} as well as by 
Jackiw and Kosteleck\'{y} \cite{jk}, in which the
$b_\m$-linear contribution to the induced Chern-Simons term was calculated. 
We understand the identity of the lowest-order calculation with
the all-order calculation by following an argument of Coleman 
and Glashow \cite{cg}, (communicated to us by S. Coleman via R. Jackiw).
Consider expanding the
$b$-dependent vacuum polarization amplitude in powers of $b$ 
($b$-perturbation theory). In $n$th order there is a two-index, 
i.e. two-photon, amplitude, with $n$ chiral insertions (of $\bs \g_5$). 
All except the first order are free of linear divergences. 
(Abelian axial anomalies come only from the
triangle graph! \cite{gj}) Hence there is no ambiguity in evaluating the higher
order graphs, and the Coleman-Glashow ``trick" may be used: momentarily
let each of the two photons carry different momenta, say $p_1$ and $p_2$ 
(this means that the chiral insertions carry non-zero momentum); 
from gauge
invariance (transversality) in {\em each} of the photons, we learn that
the amplitude is $O(p_1)$ and $O(p_2)$; i.e. it is $O(p_1 p_2)$; 
now go to equal momenta,
$p_1=p_2=p$, and observe that the amplitude must be $O(p^2)$. The
Chern-Simons term one is seeking is $O(p)$; hence all these higher-order
graphs do not contribute. The only contribution comes from the lowest
order, which is regularization dependent (or unique if the method of
evaluation by Jackiw and Kostelecky \cite{jk} is adopted).

~\\


\acknowledgements
I thank Professor R. Jackiw for enlightening
discussions and acknowledge the Center for Theoretical 
Physics, MIT, for the warm hospitality. This work was supported in part by 
the Korea Science and Engineering Foundation, 
and in part by the United States Department of Energy under grant number 
DF-FC02-94ER40818.


\begin{thebibliography}{99}                 
\bibitem{ck}                  
D. Colladay and V.A. Kosteleck\'{y}, Phys. Rev. D {\bf 58}, 116002 (1998).
\bibitem{djt}                  
S. Deser, R. Jackiw, and S. Templeton, Ann. Phys. (NY) {\bf 140}, 372 (1982);
(E) {\bf 185}, 406 (1988).
\bibitem{cfj} 
S. Carroll, G. Field, and R. Jackiw, Phys. Rev. D {\bf 41}, 1231 (1990).
\bibitem{co}
J.-M. Chung and Phillial Oh, hep-th/9812132.
\bibitem{jk} 
R. Jackiw, V.A. Kosteleck\'{y}, Phys. Rev. Lett. {\bf 82}, 3572 (1999).
\bibitem{ch}
W.F. Chen, hep-th/9903258.
\bibitem{jm}
J.-M. Chung, hep-th/9904037.
\bibitem{fg}
C.D. Fosco and J.C. Le Guillou, hep-th/9904138.
\bibitem{pv}
M. P\'{e}rez-Victora, hep-th/9905061.
\bibitem{cg}
S. Coleman and S.L. Glashow,  Phys. Rev. D {\bf 59}, 116008 (1999).
\bibitem{gj} 
I. Gerstein and R. Jackiw, Phys. Rev. {\bf 181}, 1955 (1969).  
\end{thebibliography}
\end{document}